\documentclass[a4paper]{spie}
\usepackage{graphicx}
\usepackage{latexsym}
\usepackage{cite}
\usepackage{color}

\title{CEA Bolometer Arrays: the First Year in Space}

\author{Nicolas Billot\supit{a}, M. Sauvage\supit{b}, L. Rodriguez\supit{b}, B. Horeau\supit{b}, C. Kiss\supit{c}, H. Aussel\supit{b}, K. Okumura\supit{b}, O. Boulade\supit{b}, B. Altieri\supit{d}, A. Poglitsch\supit{e}, P. Agn\`{e}se\supit{f}
\skiplinehalf \supit{a}NASA Herschel Science Center, 
Caltech, Pasadena, 91125 CA, USA \\ \supit{b}CEA, Laboratoire AIM, Irfu/SAp, Orme des Merisiers, F-91191
Gif-sur-Yvette, France \\ \supit{c}Konkoly Observatory of the Hungarian Academy of Sciences, Budapest, Hungary \\ \supit{d}Herschel Science Centre, European Space Astronomy Centre, ESA, Spain \\ \supit{e}Max-Planck-Institut f\"{u}r extraterrestrische Physik, 85748 Garching, Germany \\ \supit{f}CEA-LETI, Minatec 17 rue des Martyrs, 38054 Grenoble Cedex 9, France}

\authorinfo{Further author information: Nicolas Billot, e-mail:
nbillot@ipac.caltech.edu}

\begin{document} 
 \maketitle 

\begin{abstract}

The CEA/LETI and CEA/SAp started the development of far-infrared filled bolometer arrays for space applications over a decade ago. The unique design of these detectors makes possible the assembling of large focal planes comprising thousands of bolometers running at 300 mK with very low power dissipation. 
Ten arrays of 16x16 pixels were thoroughly tested on the ground, and integrated in the Herschel/PACS instrument before launch in May 2009. These detectors have been successfully commissioned and are now operating in their nominal environment at the second Lagrangian point of the Earth-Sun system.
In this paper we briefly explain the functioning of CEA bolometer arrays, and we present the properties of the detectors focusing on their noise characteristics, the effect of cosmic rays on the signal, the repeatability of the measurements, and the stability of the system. 
\end{abstract}

\keywords{Bolometer arrays, Far Infrared, Cold readout multiplexing, Filled arrays, Herschel/PACS.}

\section{INTRODUCTION}
\label{sect:intro}

In the 90's, the far infrared and submillimeter astronomical community already had access to sensitive detectors and background limited instruments\cite{bock,wang,kreysa,holland} . These instruments contained a modest number of pixels in the focal plane making them relatively inefficient at mapping the sky. The challenge for subsequent technological improvements was therefore to increase significantly the number of bolometers in the focal plane and build high mapping efficiency instruments. This was the main motivation for CEA to develop bolometer arrays to fly aboard the Herschel Space Observatory\cite{pilbratt} . The technological breakthrough came when CEA/LETI introduced the concept of the resonant $\lambda/4$ cavity\cite{agnese99} to optimize the absorption efficiency while dispensing with traditional bulky feedhorns. In 2000, CEA joined the Photodetector Array Camera and Spectrometer\cite{poglitsch} (PACS) consortium and equipped the PACS Photometer\cite{billot2006} with 10 filled bolometer arrays.
On May 14th 2009, the Herschel Space Observatory was launched by an Ariane-5 from Kourou, French Guyana, and it was placed in orbit around the second Lagrangian point L2 of the Sun-Earth system. Since then, the PACS Photometer has successfully passed the commissioning phase, the performance verification phase, and it entered the routine phase of the observatory in December 2009. 

Over its first year of operation in space, the PACS Photometer has been heavily demanded by the astronomical community: the PACS bolometer arrays have actually been used over 71\% of the total Herschel science observing time\footnote{This includes the observing time of PACS when used as a prime instrument or in the SPIRE-PACS Parallel mode.}. This is especially remarkable considering that the PACS Photometer is only 1 out of 5 sub-instruments on the observatory (including SPIRE and PACS spectrometers as sub-instruments).

In this paper, we first briefly describe the architecture of CEA bolometer arrays and the PACS Photometer Focal Plane Unit in section~\ref{sect:bolo}. Then we present the information, extracted from one year of observations, which is relevant to monitoring the performances and behavior of the detectors such as noise properties, stability, non-linearity, cosmic rays, and cross-talk in section~\ref{sect:inSpace}.

\section{CEA bolometer arrays}
\label{sect:bolo}

\subsection{Filled Multiplexed Bolometer Arrays}
\label{subsec:bolo}

The design of CEA bolometer arrays was driven by a requirement of high mapping efficiency. Griffin et al. 2002\cite{griffin} have shown that a bolometer array filled with bare pixels can in principle map the sky up to 3.5~times faster than bolometer arrays using feedhorns to couple the telescope with the detectors. Dispensing with feedhorns and filling the focal plane with bare pixels requires however a significant increase in the total number of bolometers operating at sub-kelvin temperature while keeping the same field of view. This has two main consequences on the architecture of the bolometer arrays:
\begin{itemize}
\item It is necessary to multiplex the readout electronics at 300~mK to keep the power dissipation at this temperature stage within the limited budget of $\sim$10~$\mu$W. At the time, even though CMOS transistors are relatively noisy ($\sim$3~$\mu V / \sqrt{Hz}$) compared to the JFET transistors traditionally used to read germanium thermometers\cite{turner} , it was the only technological solution to implement the time-domain 16-to-1 multiplexing scheme. Then in order to reach the sensitivity requirements, bolometers had to have very high responsivities to overcome the \emph{noisy} CMOS electronics, hence a very high impedance of the order of $10^{12}$~$\Omega$.
\item CEA/LETI opted for an all-Silicon design to manufacture monolithic bolometer arrays in a collective manner while exploiting mature micromachining techniques and reducing development costs. Thermometers are also made of Silicon (with Phosphorus implantation compensated with Boron to achieve the required high impedance).
\end{itemize}
The first operational $16\times 16$ bolometer array was manufactured in 1999. A detailed description of CEA bolometer arrays can be found in Agn\`{e}se et al. 2003\cite{agnese} and Billot et al. 2006\cite{billot2006} .

\subsection{The PACS Photometer Focal Plane Unit}
\label{subsec:phfpu}

The PACS Photometer Focal Plane Unit is a dual-band imaging camera. It contains two Bolometer Focal Planes (BFP) observing simultaneously the same field of view of 3.5$\times$1.75~arcmin$^2$. On each side of the PACS Photometer field of view\cite{poglitsch} are internal calibration sources used for monitoring the detectors gain variations (see section~\ref{subsec:stabili}). The short (long) wavelength channel, also called Blue (Red) channel, operates from 60 to 130~$\mu$m (130 to 210~$\mu$m). The Blue BFP contains 8~$16\times 16$ bolometer arrays butted together, and the Red BFP contains 2~bolometer arrays. A detailed description of the PACS Photometer can be found in Billot et al. 2006\cite{billot2006} . The native sampling frequency of the arrays is 40~Hz, however due to the limited downlink bandwidth of the Herschel spacecraft, the signal has to be processed on-board leading to an effective sampling frequency of 10~Hz.

\subsection{Cryo-cooler}
\label{subsec:cryo}

CEA bolometer arrays operate at $\sim$300~mK. A $^3$He sorption cooler provides a base temperature of 285~mK to the focal plane for the whole duration of a PACS Photometer observation campaign. A more detailed description of the cryo-cooler is given in Billot et al. 2006\cite{billot2006} and in Duband et al. 2008\cite{duband2008} . \\
Each instrument campaign starts with a cryo-cooler recycling that lasts 3~hours. The evaporator temperature ($T_{EV}$) reached after a cooler recycling is $285.5\pm1.5$~mK. During the first year in space, 72~cryo-cooler recyclings have been executed. The hold time of the cryo-cooler depends on the energy dissipated on the focal plane. Most of the energy dissipated on the focal plane comes from operating the bolometer arrays. The hold time therefore depends on the detectors operating time in a very predictable way\cite{duband2010} . The usual hold time is around 64~hours, but it can reach 73~hours when the bolometers are little used.

\section{One year in Space}
\label{sect:inSpace}

\subsection{Bolometers Sensitivity}
\label{subsec:perf}

Ground measurements of the bolometer arrays performances consisted in staring observations of a warm black body to estimate the noise properties (in units of $V/\sqrt{Hz}$), and in modulated observations between two black bodies at different temperatures to measure the responsivity of the bolometers (in units of $V/W$). The sensitivity of the detectors, or Noise Equivalent Power (NEP, in units of $V/W$), was derived as the ratio of the noise to the responsivity. This parameter was optimized on the ground with respect to bolometer bias settings and the optical load. The results are presented in Billot et al. 2006\cite{billot2006}.\\
A similar approach for measuring PACS performances in-flight was not possible since the only black bodies available are the PACS internal calibration sources which exhibit prohibitively long relaxation times (consequently very stable temperatures). So we made use of celestial primary calibrators\footnote{A primary calibrator is a celestial object for which we have highly reliable models of its FIR emission.} to measure the noise and responsivity for the actual telescope foreground illumination. Besides it is more relevant to estimate the camera sensitivity from celestial observations that include the whole chain of detection rather than using internal sources. Presently, we find NEPs of $2.5\times10^{-16}$~$W/\sqrt{Hz}$ for the Blue channel arrays and $2.8\times10^{-16}$~$W/\sqrt{Hz}$ for the Red channel arrays. These values are consistent with previous ground measurements.

\subsection{Glitches}
\label{subsec:glitch}

The Herschel Space Observatory orbits around the second Lagragian point L2. The space environment at this location consists of protons ($\sim$80\%), alpha particles ($\sim$14\%), and heavy ions ($\sim$4-5\%) originating from galactic cosmic rays, solar events, and solar wind plasmas. The shielding around the PACS instrument can be represented as a 11~mm thick aluminum sphere, which corresponds to 3~g/cm$^2$ of aluminum. Protons and alpha particles with energy above 90~MeV and 200~MeV, respectively, can penetrate such a shield. Considering the cosmic particle spectra derived by the OMERE\cite{peyrard} software at L2, we expect a total flux of 0.29~particles/cm$^2$/s/sr on the PACS detectors (0.27~particles/cm$^2$/s/sr coming from protons, and 0.02~particles/cm$^2$/s/sr coming from alpha particles). Taking into account the geometry of an individual bolometer array ($1.2\times 1.2\times 0.04$~cm), Horeau et al. 2006\cite{horeau} derived an expected hit rate of 3~particles/s per array from primary particles. Nuclear and gamma-ray reactions in the material surrounding the detectors create secondary particles that also contribute to the total flux of particles through the array. Horeau et al. finally predicted that the glitch rate per individual PACS module at L2 would be $\sim$5~particles/s.

We exploit the observation of a distant lensing galaxy cluster to measure the actual glitch rate at L2. This data set is quite appropriate for analyzing glitch statistics since the astronomical signal is so faint that it is not distinguishable from the noise along single timelines. In addition this data set was obtained in a special engineering mode in which the signal is downlinked for 1~matrix in the Blue BFP (8~pixels in the Red BFP) at the native 40~Hz sampling frequency instead of the whole array at the nominal 10~Hz. The higher sampling frequency of this data set enables a better sampling of the glitch temporal evolution, and therefore improves their detection and characterization. 

\begin{figure}
	\begin{center}
    		\begin{tabular}{ccc}
    			\includegraphics[height=0.23\textwidth]{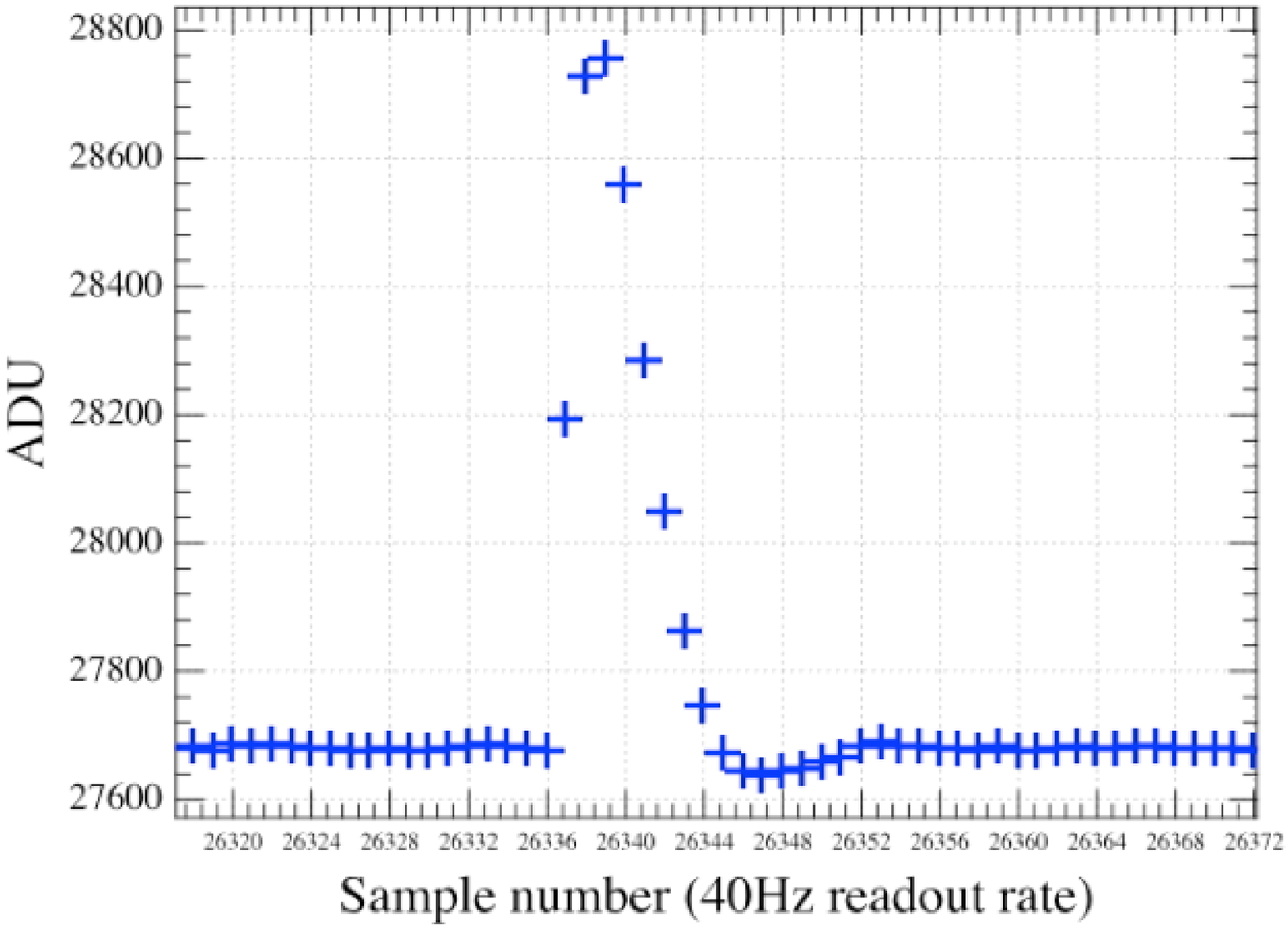} &
    			\includegraphics[height=0.23\textwidth]{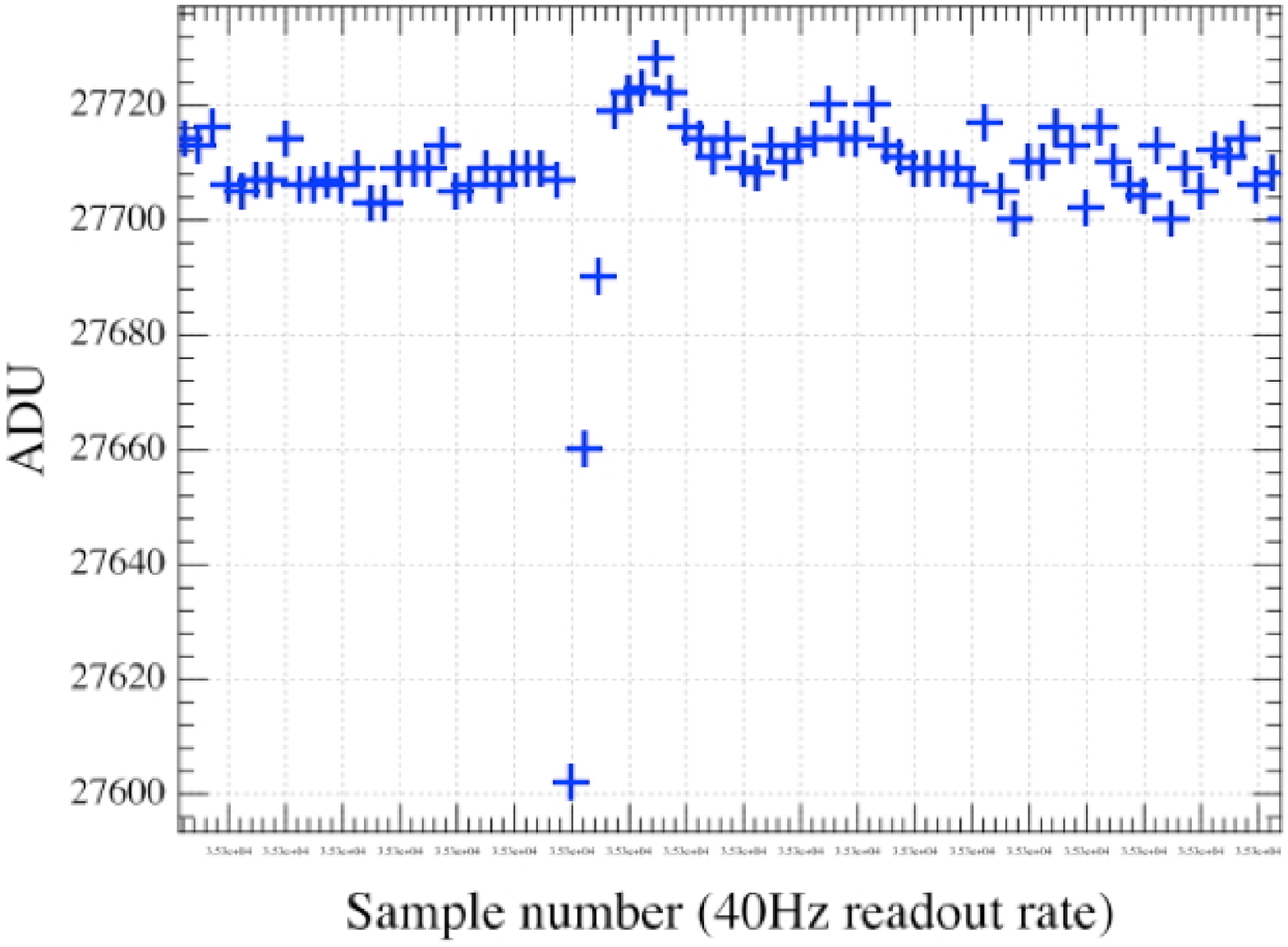} &
    			\includegraphics[height=0.23\textwidth]{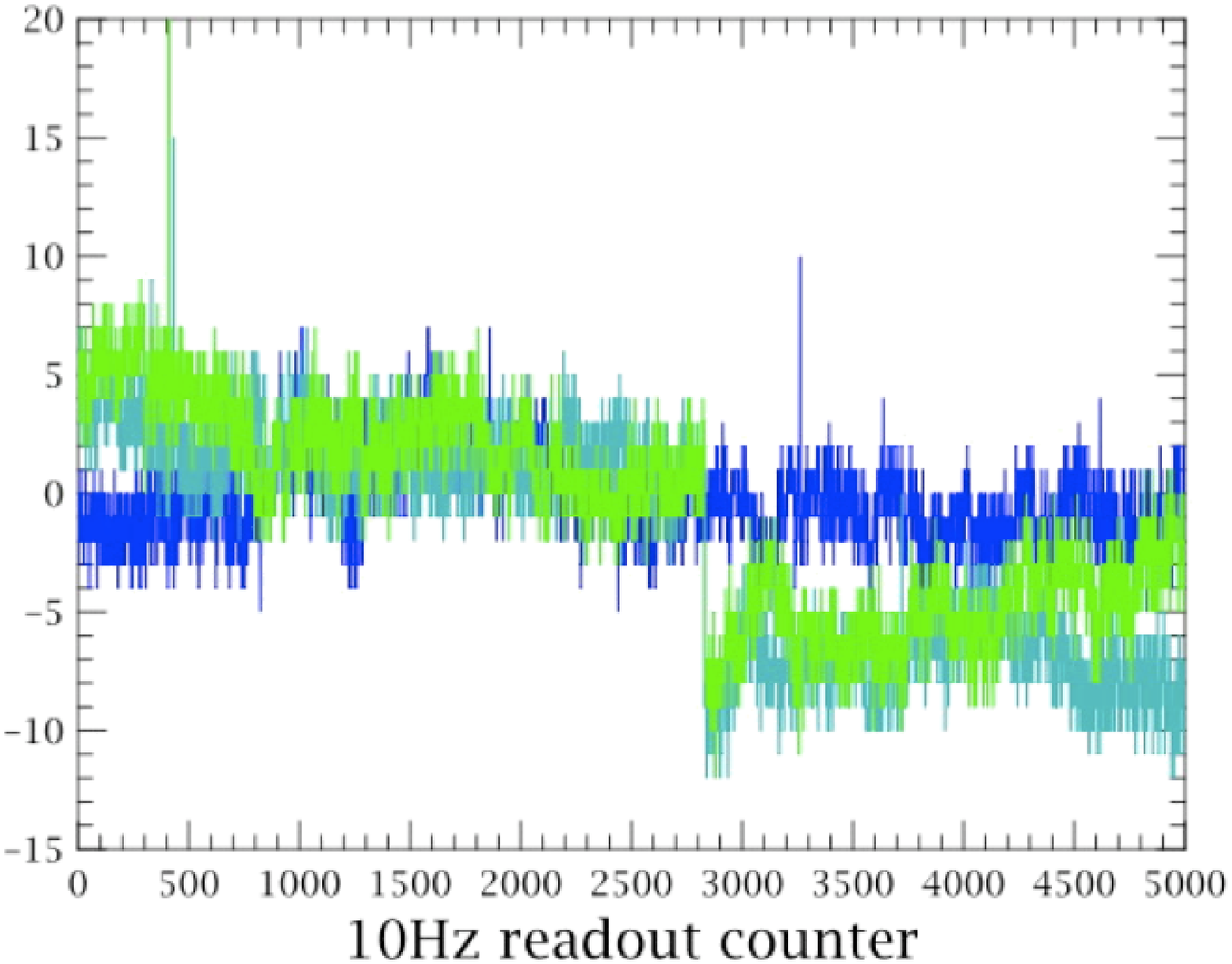}
     		\end{tabular}
    		\caption{Glitch zoology. \emph{Left panel}: Positive glitch due to an impact in the suspended grid. \emph{Middle panel}: Negative glitch due to an impact in the inter-pixel wall. \emph{Right panel}: Offset discontinuity with long recovery due to an impact in the readout electronics.}  \label{fig:glitch}
	\end{center}
\end{figure}

\begin{figure}[htbp]
	\begin{minipage}{0.43\linewidth}
		\begin{center}
			\begin{tabular}{c}
				\includegraphics[width=0.9\textwidth]{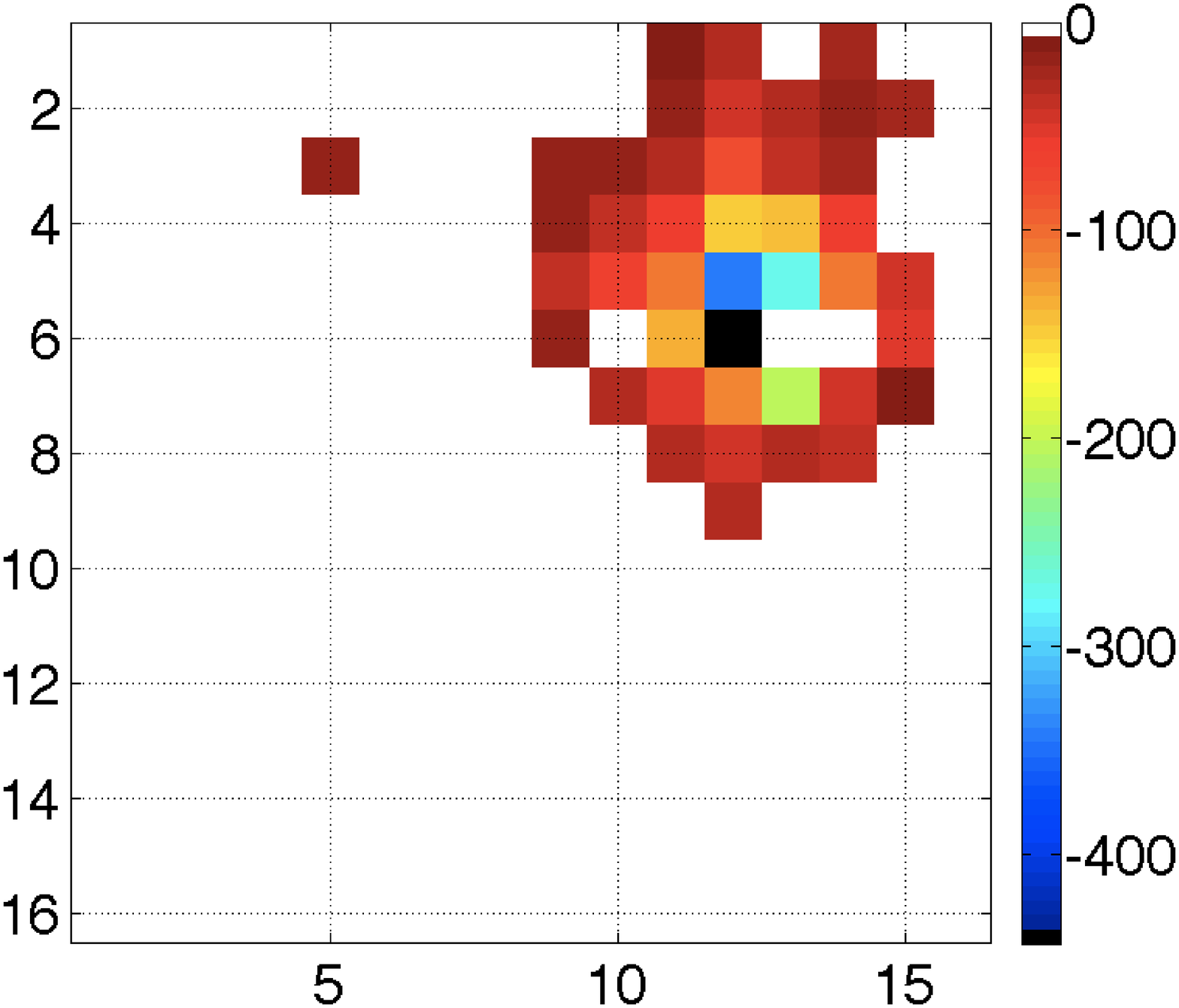}
			\end{tabular}
		\end{center}
		\caption{\label{fig:glitch_Xtalk} Thermal cross-talk between contiguous pixels on a $16\times16$ array when an energetic cosmic ray hits the inter-pixel wall. The central pixel of the  hit area shows the strongest glitch.}
	\end{minipage}
	\hfill
	\begin{minipage}{0.52\linewidth}
		\begin{center}
			\begin{tabular}{c}
				\includegraphics[width=0.99\textwidth]{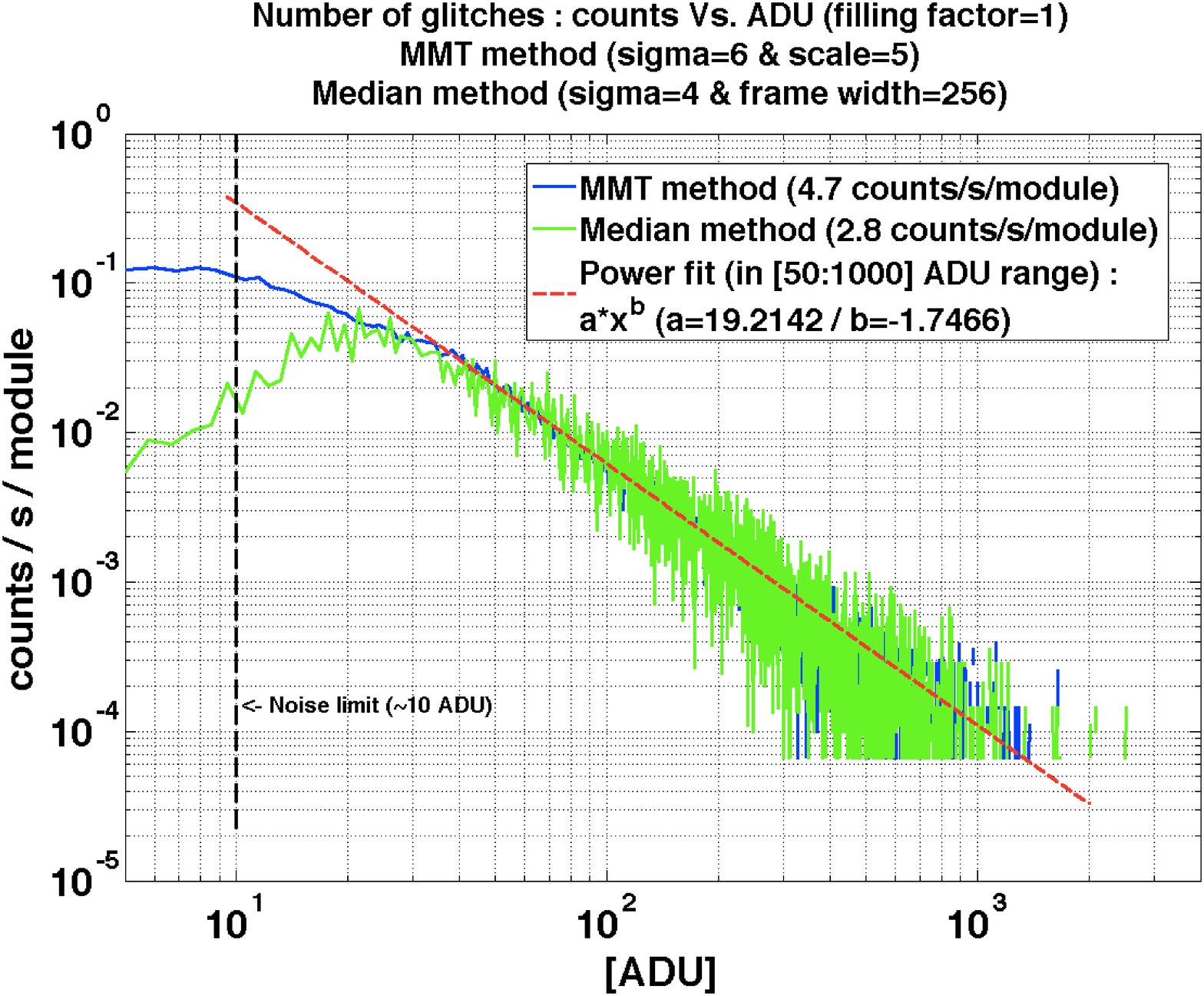}
			\end{tabular}
		\end{center}
		\caption{\label{fig:glitch_histo} Glitch number counts as a function of the glitch intensity. The two deglitching approaches are consistent down to 20~ADU.}
	\end{minipage}
\end{figure}

We find three types of glitches in the PACS Bolometer data (see figure~\ref{fig:glitch}):
\begin{itemize}
\item The most frequent events ($\sim 85$\%) are positive glitches that occur when cosmic rays hit the suspended grid of a bolometer. The energy absorbed in the grid leads to a rapid temperature rise (hence a bolometer impedance drop) and a subsequent rapid signal increase. The deposited energy is then evacuated to the heat sink through the suspending rods (see section~\ref{subsec:bolo} and references therein), which takes a few readouts to reach its previous equilibrium state. Undershoots are usually observed following the glitch, but no satisfactory physical interpretations have been found yet. 
\item Negative glitches occur when a particle hits the inter-pixel walls. Because reference thermometers are implanted on these walls, any thermal disturbances to the wall temperature would unbalance the bolometric bridge and cause the signal to drop. The amplitude of these negative signal drops is however small compared to those observed with positive glitches. This is primarily due to the large thermal capacitance $C_{wall}$ of the relatively bulky walls (compared to the suspended grids), hence the very small temperature variations $\Delta T=\frac{\Delta E}{C_{wall}}$ for a given quantity of absorbed energy $\Delta E$. Furthermore, the inter-pixel wall is the common heat sink to all the bolometers of a given matrix, a hit in the walls might consequently affect more than one pixel through a thermal cross-talk. We observe indeed that a circular region of contiguous pixels are affected by strong negative glitches (see figure~\ref{fig:glitch_Xtalk}).
\item Finally we find rare occurrences of cosmic ray hits in the readout electronics, certainly because of the small cross-section of the transistors. This type of glitch appears in the signal as a discontinuous jump in the baseline of the temporal sequence. The recovery time for these glitches is usually quite long (minutes to hours). Because the signal is multiplexed, i.e. a column of 16 pixels shares the same electronics components in the buffer unit, entire columns are generally affected by this type of glitch.
\end{itemize}
We used two algorithms to automatically detect glitches in individual bolometer time lines: (a) Median filtering of the temporal signal to remove the low-frequency noise, and sigma clipping of the flat-baseline signal to detect glitches ($4\sigma$ threshold); and (b) Multi-resolution Median Transform\cite{stark98} (MMT) which consists in decomposing the signal into different scales and look for spikes at each scale.

Figure~\ref{fig:glitch_histo} presents the histogram of positive and negative glitches detected in the 30-minutes deep field observation mentioned above. The observed distribution of glitches is a power law of the form $glitchNumber = glitchPeak^{-\alpha}$ with $\alpha= 1.75$. Note that this power law does not directly represent the energy distribution of the cosmic rays at L2, but rather their energy spectrum weighted by the energy-dependent cross-section of the bolometer array with a cosmic ray of a given energy. 
The measured power law appears to be valid above $\sim$30 independently of the algorithm used to detect the glitches. The median filtering approach shows a steep decrease in the glitch number counts for weak impacts while the MMT approach keeps picking up glitches even at the low end of the intensity scale. The departure from one approach to the other is mostly due to the parameter choice in the glitch detection process. In the present case, the MMT deglitching is quite aggressive as it tries to dig faint glitches out of the instrumental noise. Between 10 and 30~ADU, the MMT algorithm picks up faint glitches but also false glitch detections from local peaks in the instrumental gaussian noise. The median filtering approach was set to detect glitches above $4\sigma$, thus leaving faint glitches undetected. Below $\sim$10~ADU, all glitches, if any, are swamped in the sea of instrumental gaussian fluctuations, such that it becomes impossible to distinguish glitches from noise at this level.\\
Assuming a noise limit of 10~ADU, the integrated number of glitches is found to be 4.7~counts/s/module and 2.8~counts/s/module with the MMT and median filtering approach respectively. This result is consistent with the expected 5~particle hits per second per matrix derived in Horeau et al. 2006\cite{horeau} . \\
Finally, no noticeable performance degradation was observed after one year of irradiation at L2.

\subsection{Noise Properties}
\label{subsec:noise}

\subsubsection{Spectral Noise Distribution}
\label{subsubsec:noise_spec}

\begin{figure}
	\begin{center}
    		\begin{tabular}{ccc}
    			\includegraphics[width=0.45\textwidth]{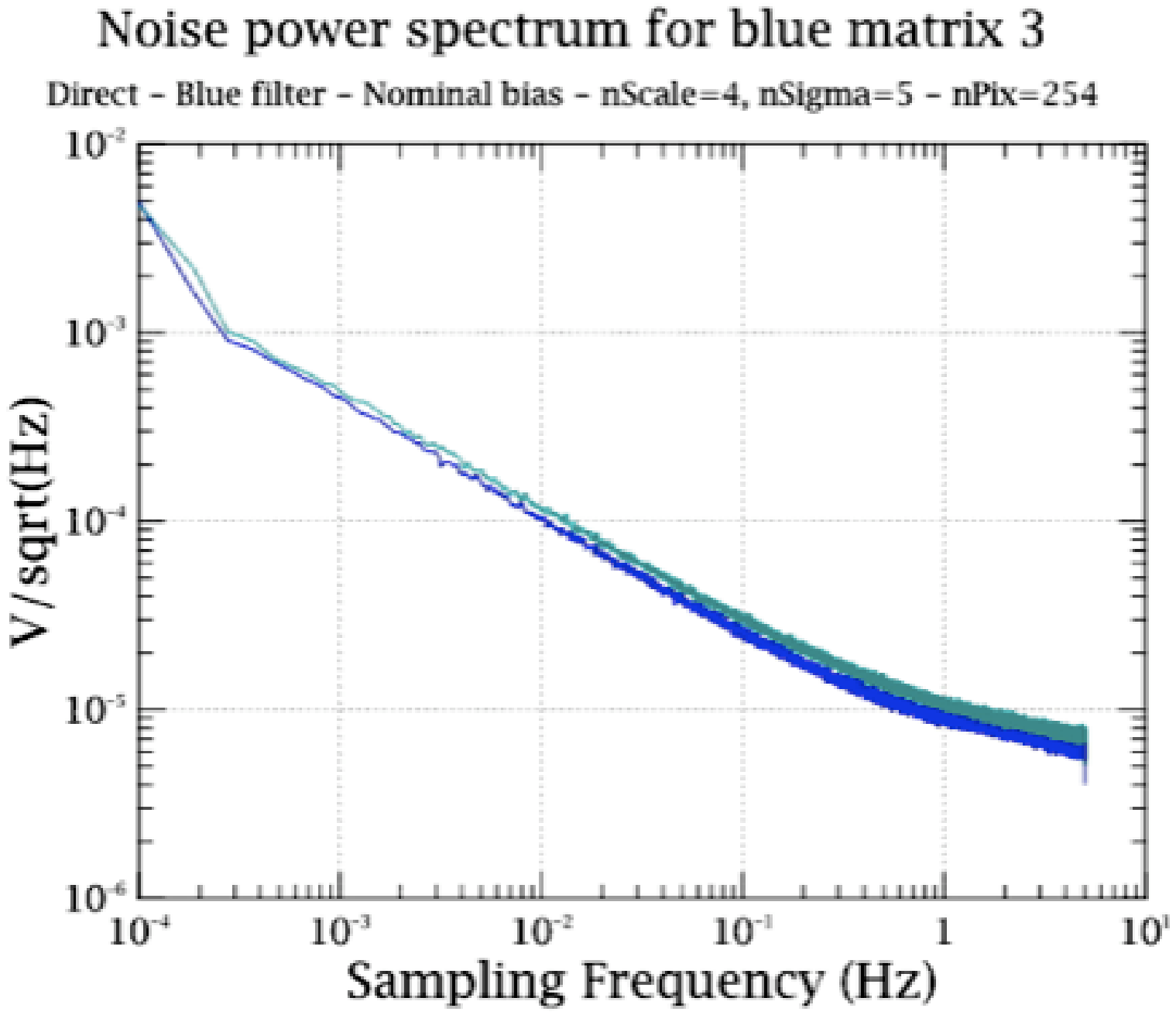} &
    			\includegraphics[width=0.45\textwidth]{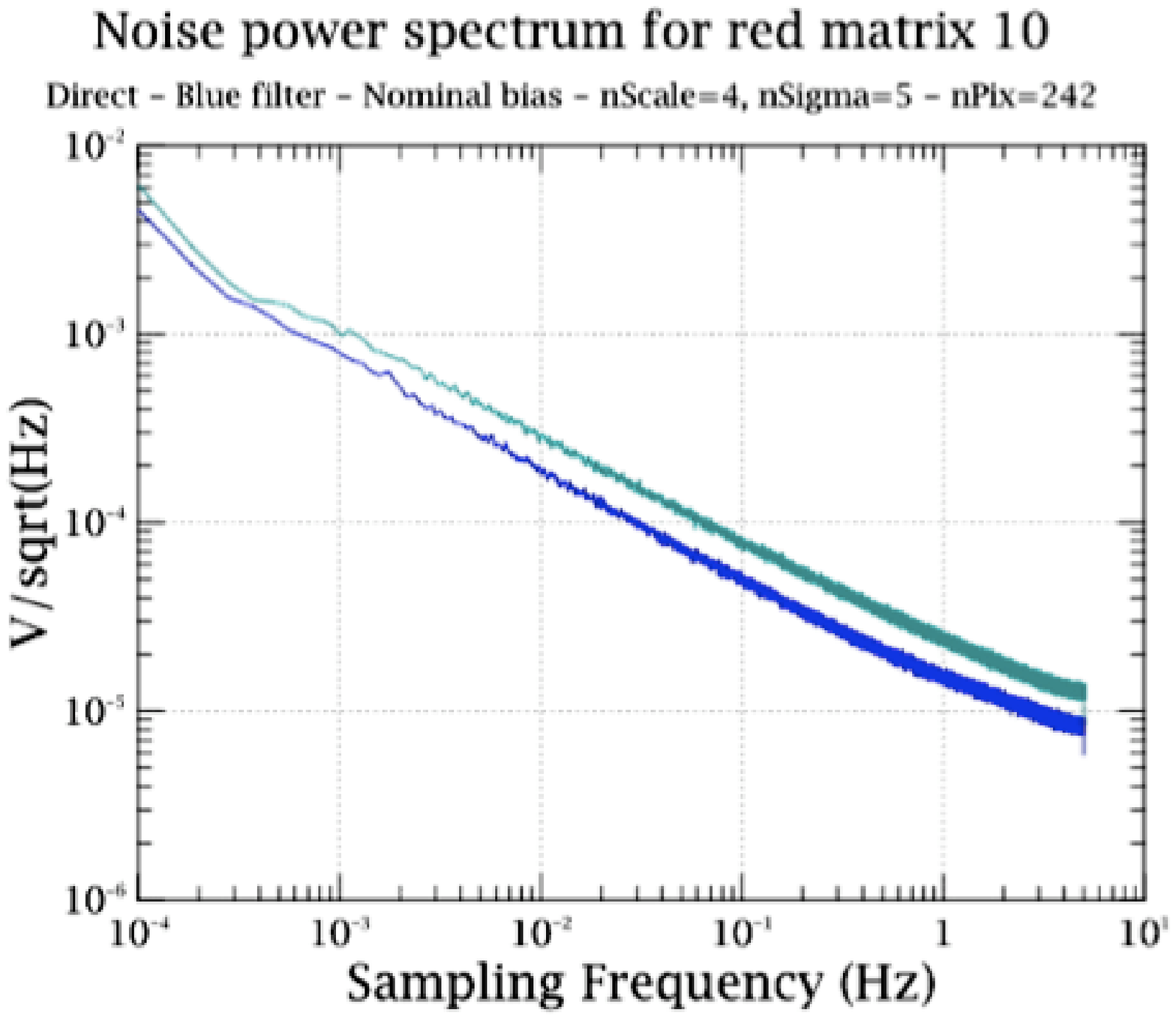} 
     		\end{tabular}
    		\caption{Noise spectral distribution averaged over a whole blue (left) and red (right) matrix. Grey and blue curves represent ground and in-flight measurements respectively. The flattening of the spectrum on the blue matrix means that the white noise contribution becomes comparable to the low-frequency noise above 1~Hz. The in-flight spectra were derived from data deglitghed using the MMT method.}  \label{fig:spectrum}
	\end{center}
\end{figure}

The PACS Photometer noise spectrum has two contributions: the actual bolometers, i.e. the high-impedance end of the circuit, and the downstream readout electronics chain. The readout electronics generates a flat noise spectrum (Johnson noise) and a low-frequency noise excess at frequencies below $\sim1$~Hz. Bolometers noise spectra are very similar with a low-frequency noise excess and a white noise component visible at higher frequencies. In addition, the bolometer noise is inherently lowpass filtered by the electro-thermal time constant of the bolometers ($\sim30$~ms). The shape of the noise spectrum can therefore be decomposed into three regions: (1) a low-frequency excess noise below $\sim1$~Hz due to slow drifts in the bolometer and electronics signal offsets, (2) a bolometer-and-electronics white noise regime between 1 and 5~Hz, and (3) a white noise contribution from the electronics only above 5~Hz. Billot et al. 2006\cite{billot2006} present a noise spectrum measured at the native sampling of 40~Hz during the flight model detector selection phase, and the two inflection points separating these 3 regimes appear clearly.\\
The nominal sampling frequency in-flight is 10~Hz due to the on-board averaging, the noise spectrum is therefore sampled up to 5~Hz such that the imprint of the bolometers time constant is not visible and the spectrum is essentially a power law of the form $f^{-1/2}$ over the whole bandpass (the white noise regime is only marginally visible on the in-flight spectra). Figure~\ref{fig:spectrum} presents spectral noise densities measured on the ground and in-flight. The noise properties have not changed before and after launch, except for a slight improvement\footnote{this improvement is due to a modification of the bias setting of the red detectors readout electronics during the PACS commissioning phase.} on the red channel array. Note that these spectra were obtained by averaging the noise spectrum of all operational pixels on the matrix ($\sim256$~pixels) to represent the global noise properties of the bolometer array as well as to reduce the statistical fluctuations in the spectrum. The noise levels measured in-flight at 3~Hz are 6 and 10~$\mu V/\sqrt{Hz}$ in the Blue and Red channels respectively.


\subsubsection{Low-frequency Noise and Cryo-cooler Cycle Phases}
\label{subsubsec:noise_cooler}

\begin{figure}
	\begin{center}
    		\begin{tabular}{c}
    			\includegraphics[width=0.6\textwidth]{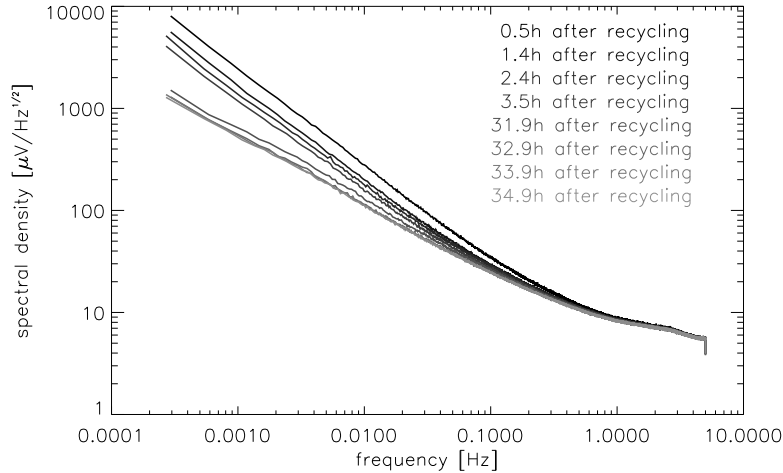} 
     		\end{tabular}
    		\caption{Noise spectral distribution measured at different phases of the cryo-cooler cycle.}  \label{fig:spectrum_cooler}
	\end{center}
\end{figure}

Bolometers are very sensitive to their operating temperature, therefore any base temperature fluctuations would be echoed in the output signal as common drifts visible for all pixels, in other words as correlated noise. The PACS cryo-cooler provides a base temperature at the evaporator that reaches 285~mK right after the cryo-cooler recycling, then it rises slowly until the end of the cryo-cooler cycle with an amplitude $\leq2$~mK. The temporal evolution of the evaporator temperature exhibits a relatively strong gradient early in the cryo-cooler cycle, and it gradually flattens out until all the $^3$He is sucked out from the evaporator\cite{duband2010} . We have measured the noise spectral densities of the bolometer arrays at different phases of the cryo-cooler cycle, and it appears that the slope of the low-frequency noise excess also evolves in time, following the evaporator temperature variations. Figure~\ref{fig:spectrum_cooler} shows that the slope of the noise spectra changes rapidly right after the cryo-cooler recycling, and seems to have reached their asymptotic value at least after $\sim30$~hours. We attribute this slope evolution to the evaporator temperature relaxation after recycling.\\
Curves from figure~\ref{fig:spectrum_cooler} might be misleading however, as it seems the noise properties are not stationary. One should keep in mind that the extra low-frequency noise excess visible after the cryo-cooler recycling is purely correlated noise, and that once the common thermal drifts are removed from the signal the stationary property of the noise is restored to individual bolometers.

\subsection{Detector Stability}
\label{subsec:stabili}

Detector stability is an essential property that ultimately sets the instrument photometric accuracy. Space instruments are in general very stable due to the constant environment in which they evolve, especially at L2. After one year in space, we have gathered enough statistics to assess the stability of the PACS bolometer arrays over long periods of time and look for possible systematic trends. We have monitored the signal and responsivity of the detectors with two independent approaches exploiting internal calibration sources and observations of celestial standards.

\subsubsection{Calibration blocks monitoring}
\label{subsubsec:calBlock}

Each PACS Photometer observation is preceded by a short calibration block. The purpose of these \emph{calBlocks} is to measure the current gain of the bolometers to be used for potential post-processing responsivity drift corrections. CalBlocks take about 30~seconds to execute and consist in a chopped observation between the two internal calibration sources (19 chopper cycles at 0.625~Hz) located on each side of the instrument field of view (see section~\ref{sect:bolo}). 

\begin{figure}
	\begin{center}
    		\begin{tabular}{ccc}
    			\includegraphics[width=0.3\textwidth]{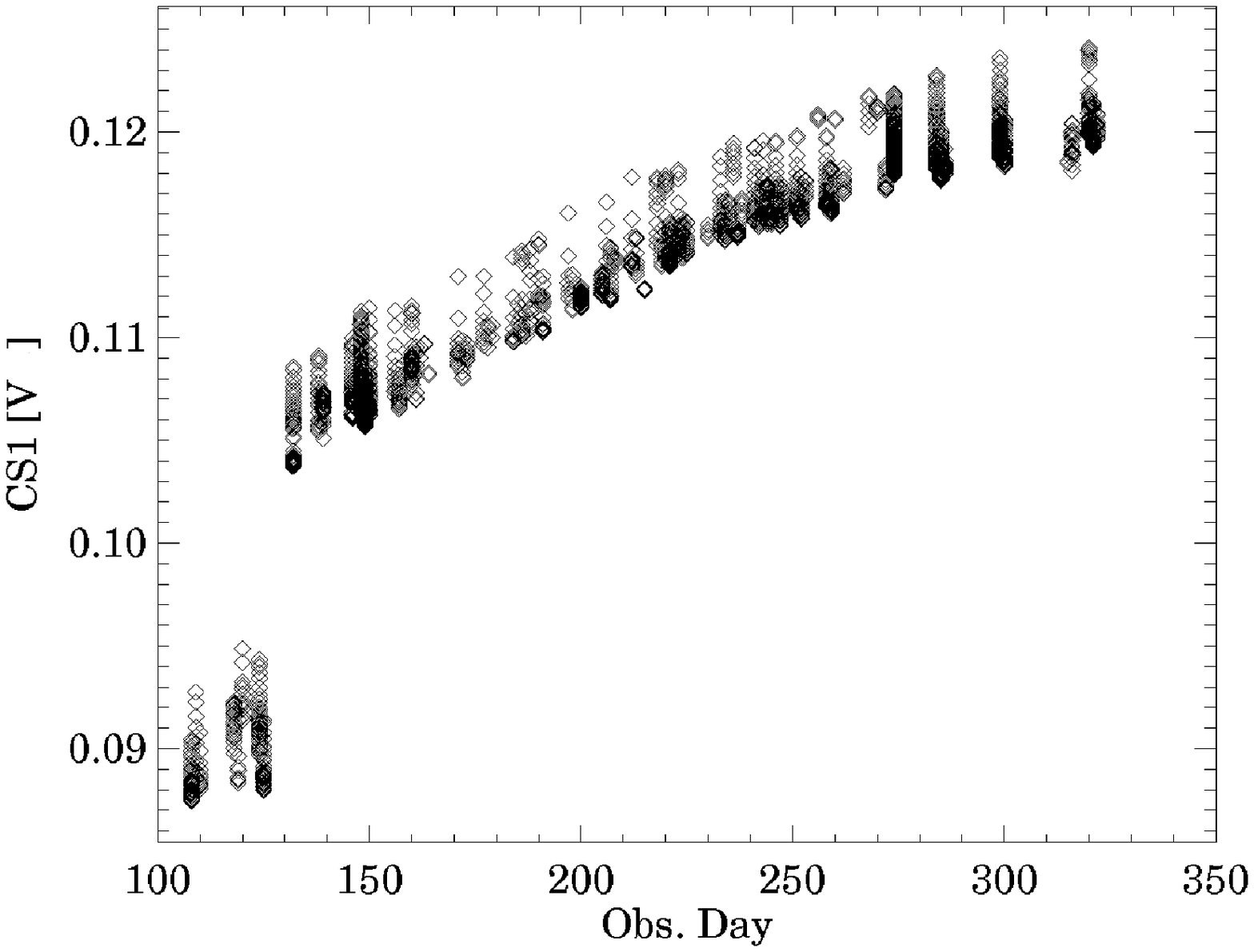} &
    			\includegraphics[width=0.3\textwidth]{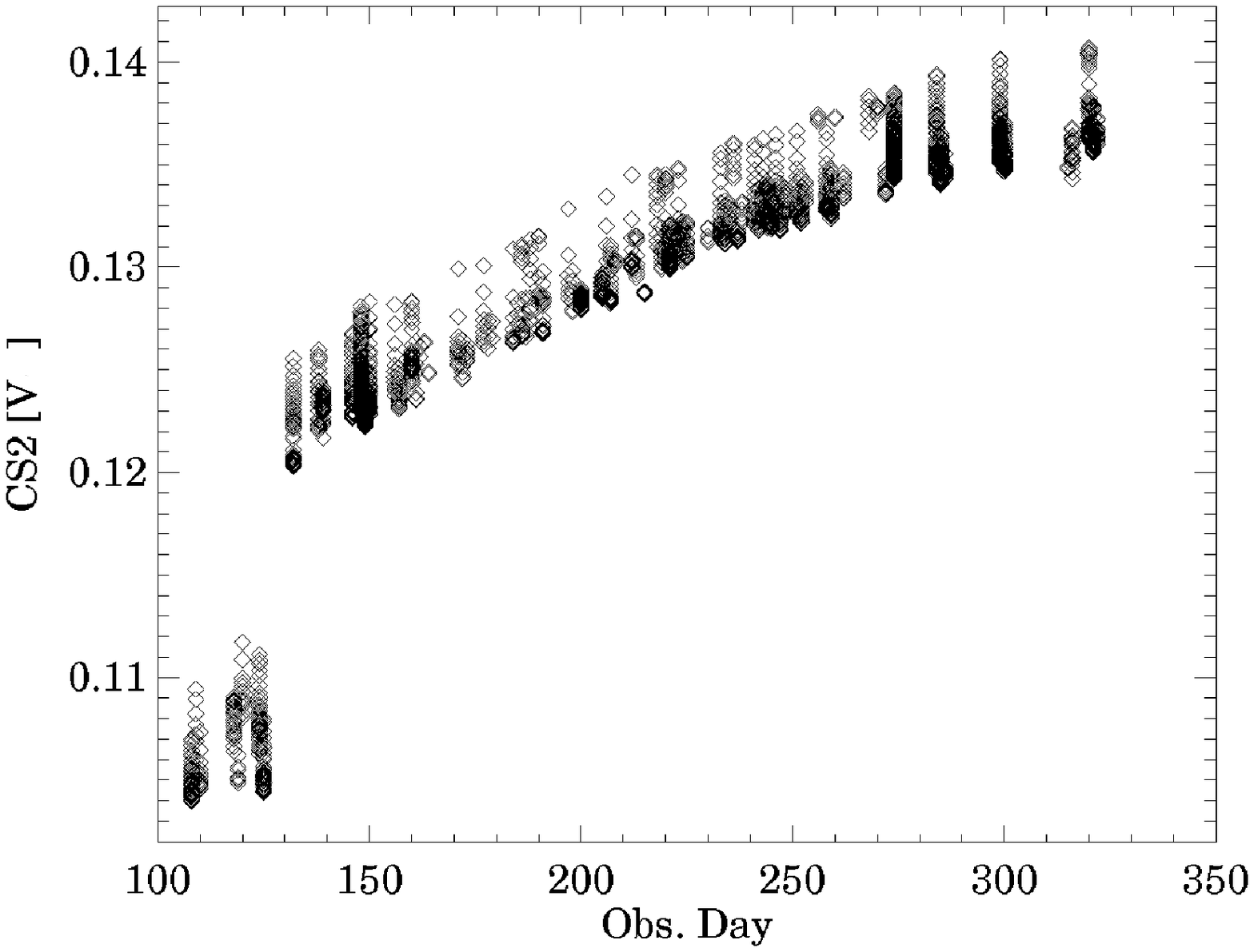} &
    			\includegraphics[width=0.3\textwidth]{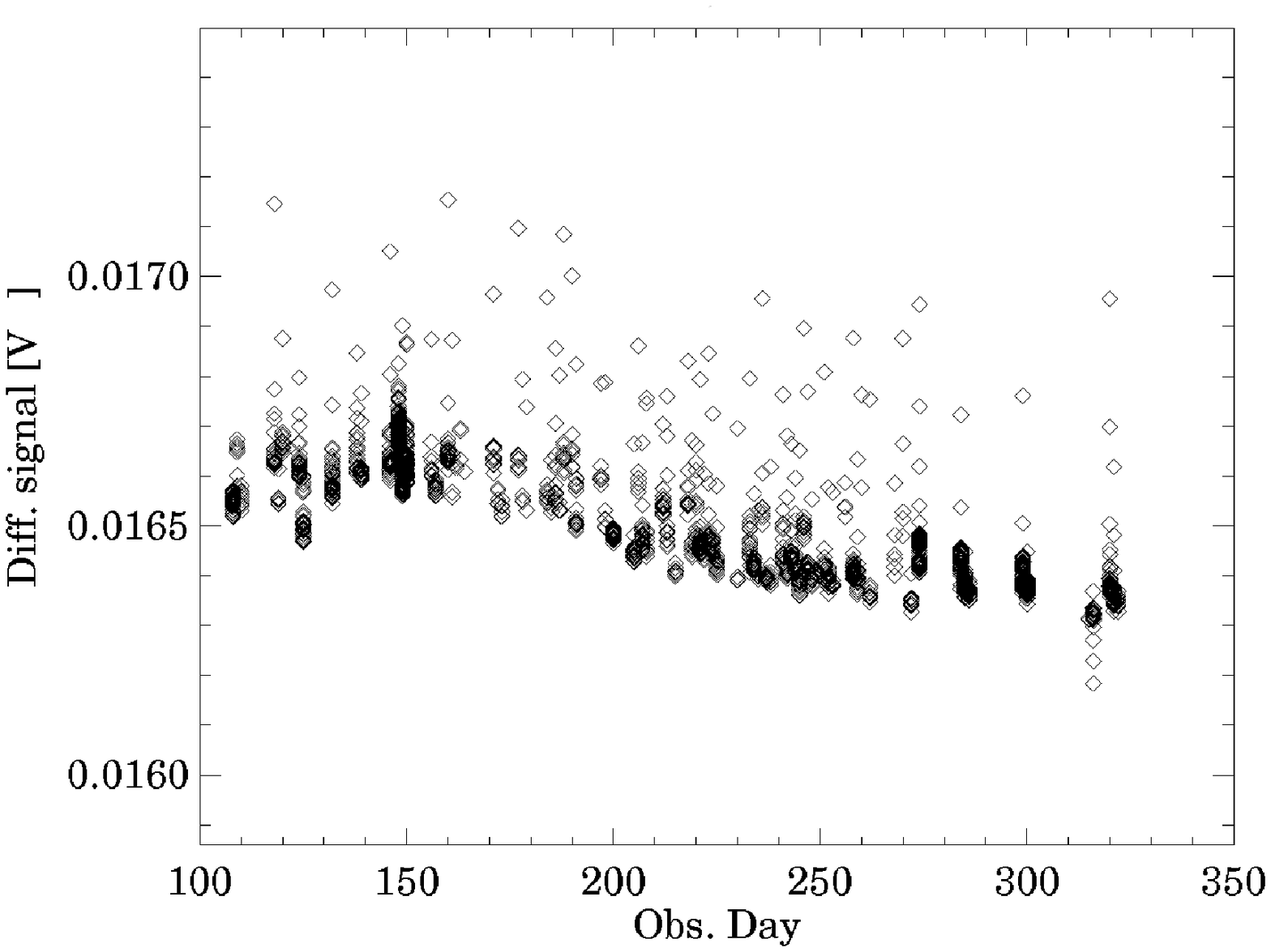} 
     		\end{tabular}
    		\caption{Evolution of the averaged signal on CS1 (left), CS2 (middle) and averaged responsivity CS2-CS1 (right) over a period of 220~days extracted from calibration block observations. Responsivity variations are within a percent.}  \label{fig:allOD}
	\end{center}
\end{figure}

\begin{figure}
	\begin{center}
    		\begin{tabular}{c}
    			\includegraphics[width=0.6\textwidth]{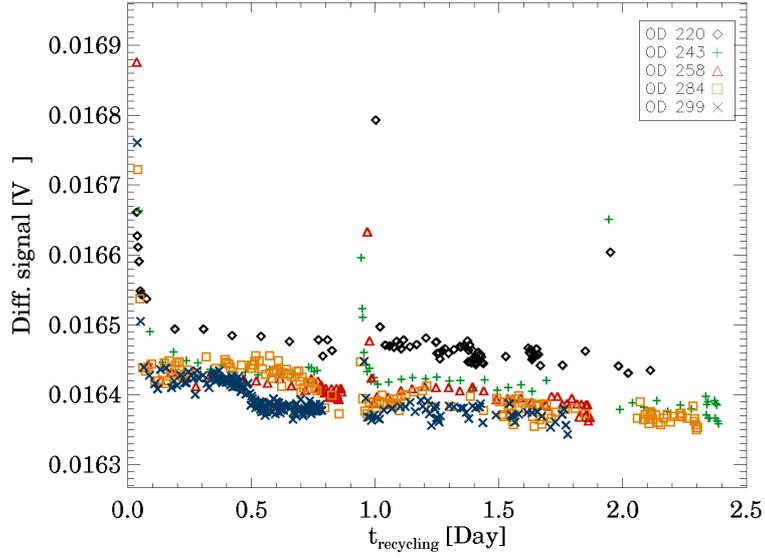}
     		\end{tabular}
    		\caption{Evolution of the bolometers responsivity as a function of the cryo-cooler cycle phase.}  \label{fig:respSomeODs}
	\end{center}
\end{figure}

Figure~\ref{fig:allOD} presents the averaged bolometer signal measured when illuminated by the two calibration sources CS1 and CS2 over a period of 220~Operational Days (OD), as well as the signal difference ($CS2-CS1$), in other words the multiplicative part of the signal, which scales as the bolometer responsivity. The latter data set (right panel) shows a long term variation on a sub-percent scale which demonstrates the extreme stability of the bolometer arrays over long periods of time. On shorter time scales, we attribute the upward dispersion of the responsivity measurements to the transient behavior of the detectors after switch-on (see also figure~\ref{fig:respSomeODs} for a subset of instrument campaigns).\\
During the commissioning phase of the PACS Photometer, we modified the detectors bias voltages to adjust the signal offset relative to the ADC dynamic range and minimize saturation events. This bias change appears as a discontinuity in the monotonic ascent of the signal observed at OD~128 in the two left panels of figure~\ref{fig:allOD}. We attribute the global trend of the signal to increase with time to the straylight coming from the primary mirror into the internal PACS calibration source signal. The Herschel primary mirror indeed shows seasonal temperature variations, with amplitude of 5-10~K, which correlates strongly with the signal long term variations seen in figure~\ref{fig:allOD}. A preliminary analysis shows that we can exploit this correlation to correct for the primary mirror straylight in the calBlocks and reduce even further the long term variations of the signal and responsivity measurements.\\

Figure~\ref{fig:respSomeODs} presents the evolution of the bolometer responsivity, derived from CalBlocks, averaged over a whole array during 5~instrument campaigns starting between OD220 and OD299. The x-axis of the graph represents the time elapsed since the last cryo-cooler recycling, $t_{recycling}$. For each instrument campaign the data gaps at $t_{recycling}\sim$1 and/or 2~days correspond to the \emph{Daily TeleCommunication Period}, or DTCP, during which the data accumulated during the past OD are downlinked to Earth, and telecommands to be executed on the following OD are uplinked to the spacecraft. PACS bolometers have to be switched off during DTCP, which explains the lack of data for approximately 2~hours every day. 
In addition bolometers are thermal detectors, and as such their responsivity is expected to vary with their operating temperature. The evaporator temperature increases indeed monotonically during a cryo-cooler cycle\cite{duband2010}  which directly influences the bolometers responsivity. This explains the tight correlation we find between the bolometer responsivity and the time after recycling, $t_{recycling}$, or equivalently with the evaporator temperature $T_{EV}$.
The data set of figure~\ref{fig:respSomeODs} also exhibits spikes at the beginning of each OD ($t_{recycling}=0,1,2$) which are due to a $\sim20$~minutes relaxation time after detectors switch on.



\subsubsection{Flux repeatability from celestial standards}
\label{subsubsec:stars}

\begin{figure}
	\begin{center}
    		\begin{tabular}{c}
    			\includegraphics[width=0.7\textwidth]{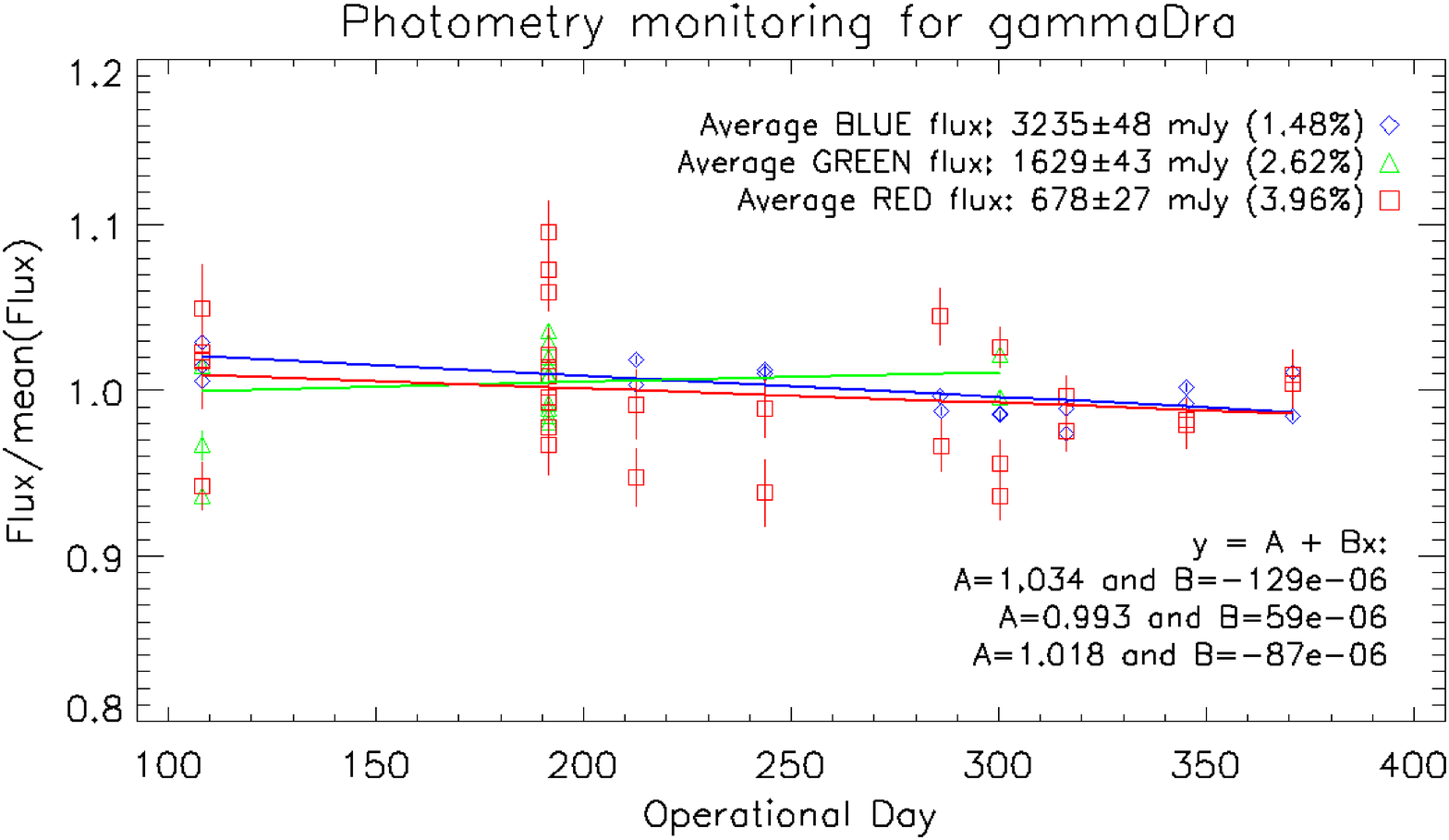} 
     		\end{tabular}
    		\caption{Photometry monitoring since the start of the Performance Verification Phase on the star $\gamma\!\!$~Draconis in scan map mode. The standard deviation from the averaged flux is reported on the top right of the plot as a percentage.}  \label{fig:gammaDra}
	\end{center}
\end{figure}

One aspect of the PACS Photometer calibration plan is to observe celestial standards on a regular basis to monitor the evolution of the detectors performances. After one year of operation, the most visited object is the star $\gamma\!\!$~Draconis. It is actually one of the dozen primary standards from which we derive the absolute flux accuracy of the instrument. The main difference of monitoring celestial standards compared to internal calibration sources is that it includes the whole chain of detection and analysis, from the first reflection off the primary mirror to the final stage of measuring the photometry of the star. We therefore expect a somewhat larger uncertainty on these flux repeatability measurements compared to the gain stability figures presented previously.\\
We have reduced 36~scan map observations of $\gamma\!\!$~Dra using a custom version of the PACS pipeline, and we have measured the source flux within relatively large apertures (20'' radius in the Blue and Green bands, and 25'' in the Red band). Figure~\ref{fig:gammaDra} presents the evolution of the flux measured for this object as a function of the observation day. The same processing and aperture photometry parameters were consistently used on all 36 data sets. It appears that flux measurements on $\gamma\!\!$~Dra are reproducible to within a few percents which confirms the very good stability performances of the detectors. We find however a weak trend in the data set that is consistent with the seasonal responsivity decrease evidenced in the right panel of figure~\ref{fig:allOD}

\subsection{Non-linearity}
\label{subsec:linear}

\begin{figure}[htbp]
	\begin{minipage}{0.48\linewidth}
		\begin{center}
	    		\begin{tabular}{c}
    				\includegraphics[width=0.95\textwidth]{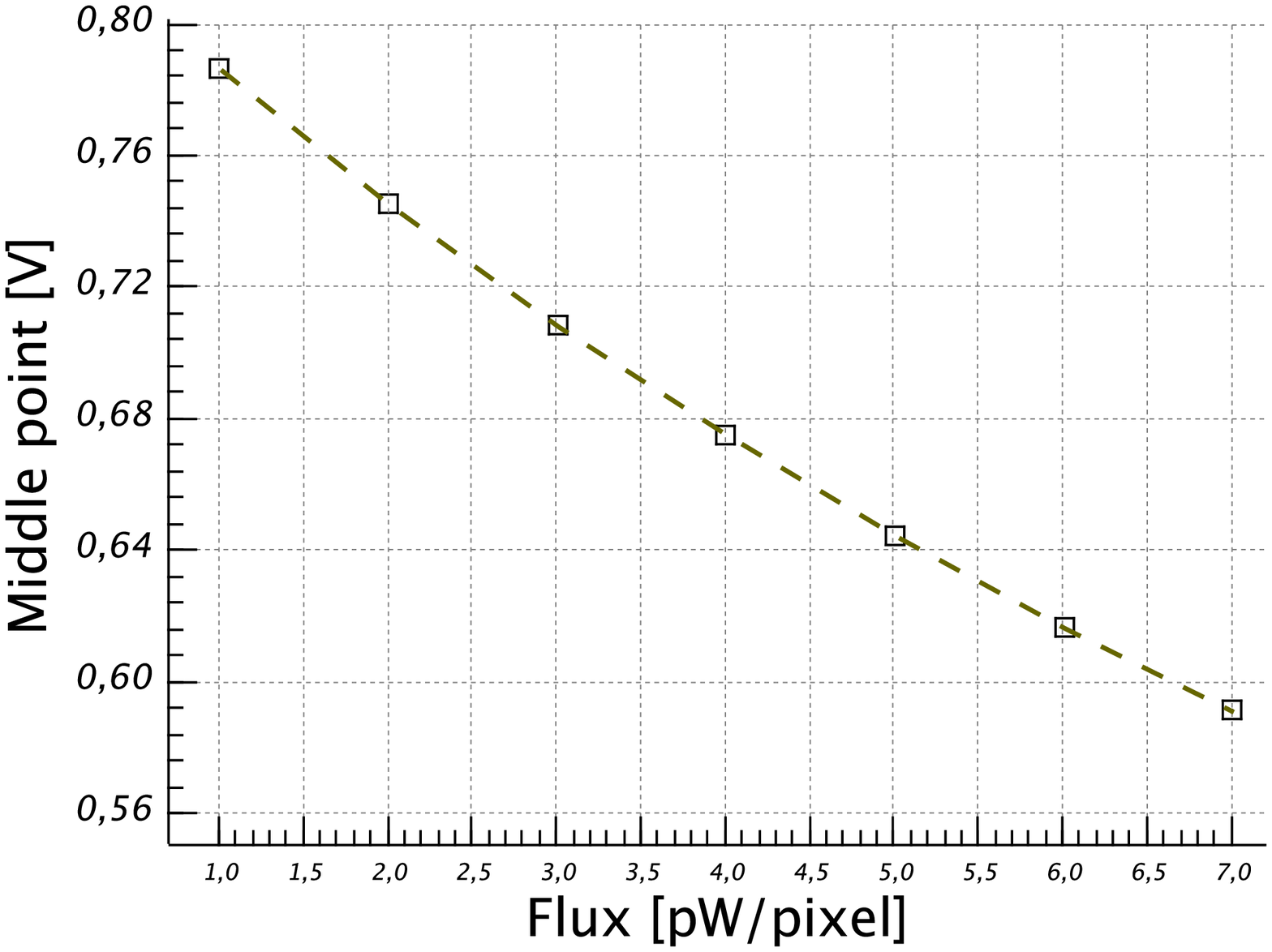} 
     			\end{tabular}
    			\caption{Reconstructed evolution of the voltage across the bolometer (middle point) as a function of the illumination. Note that the flux of most astronomical sources represents only a small fraction of a pW/pixel.}  \label{fig:nonlinear}
		\end{center}
	\end{minipage}
	\hfill
	\begin{minipage}{0.48\linewidth}
		\begin{center}
    			\begin{tabular}{c}
    				\includegraphics[width=0.95\textwidth]{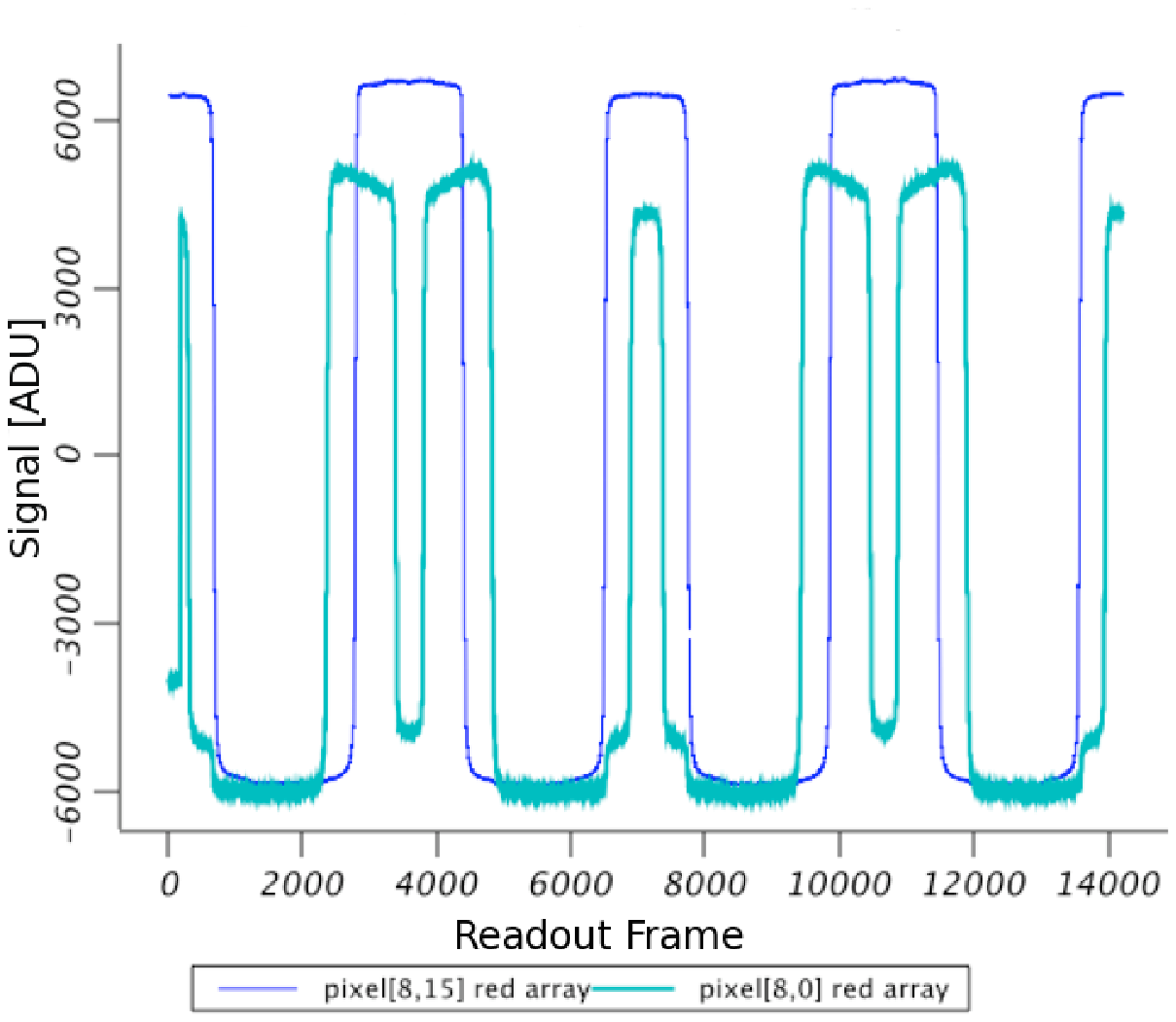} 
     			\end{tabular}
    			\caption{Cross-talk in the red array.}  \label{fig:Xtalk}
		\end{center}
	\end{minipage}
\end{figure}

Bolometers are non-linear detectors by essence. In the case of PACS, the foreground emission is very stable and the astronomical signal is a very small fraction of the primary mirror emission. The flux excursions falling on the detectors are therefore relatively small, even for bright objects. \\
Figure~\ref{fig:nonlinear} presents the evolution of the middle point of a bolometer as a function of the optical load per pixel. The middle point is the voltage measured at the middle of a bolometric bridge, i.e between the bolometer and its associated reference resistor. The middle point is actually reconstructed from the output signal of the camera while correcting for the transfer function of the whole electronics chain\cite{billot2006} . In the Blue band of the PACS Photometer, the foreground emission from the telescope represents $\sim$2~pW/pixel, and a 1~Jy source converts into $\sim$0.04~pW spread over a dozen pixels in the PSF. Figure~\ref{fig:nonlinear} shows that the curve has a large radius of curvature, hence for an operating point of 2~pW/pixel and flux excursions of less than 0.01~pW/pixel, the detectors actually operate in a quasi-linear domain. In fact, we estimate that the bolometers will depart from a linear behavior by less than a few percents for sources brighter than 50~Jy. It was actually confirmed in-flight that, within the calibration errors, the detectors responsivity is constant for flux sources up to 50~Jy.

\subsection{Cross-talk}
\label{subsec:cross}

A cross-talk event occurs when the signal from a given pixel influences the signal from another pixel. In the case of CEA bolometer arrays, electrical cross-talk may occur along readout lines because all pixels of a given line share common bits of electronics (time-domain multiplexing\cite{billot2006}). At the native 40~Hz sampling frequency, a whole line of 16~pixels is readout in 25~ms. The downstream readout electronics therefore ticks at 640~Hz. In the event of a slow pixel readout circuitry, i.e. relaxation time longer than 1.5~ms, then the remanent signal from this pixel might contribute to the readout signal from the next pixel in the readout sequence. This type of electrical cross-talk is mostly noticeable in two readout lines of the red array, but it also affects the blue array to a smaller extent. In the red array, the cross-talk occurs between pixels of line~0 and line~15, which are pixels located on opposite sides of the array. Figure~\ref{fig:Xtalk} presents an example of cross-talk observed in-flight between pixel[8,15] and pixel[8,0] of one of the red matrix. Pixel[8,0] is clearly influenced by the signal change of pixel[8,15]. 
It is possible to suppress this cross-talk by increasing the current flowing through the cold readout electronics, but that would prohibitively increase the power dissipated at the 300~mK stage. We are therefore working on the implementation of a cross-talk correction module in the PACS Photometer Pipeline.

\subsection{Interferences}
\label{subsec:interf}

\begin{figure}
	\begin{center}
    		\begin{tabular}{c}
    			\includegraphics[width=0.7\textwidth]{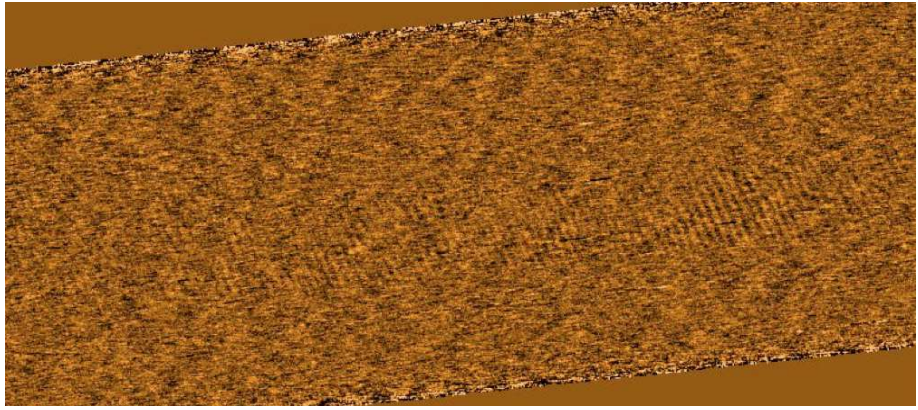} 
     		\end{tabular}
    		\caption{Interferences observed in-flight. These are presumably due to electromagnetic perturbations generated by solar panels.}  \label{fig:interf}
	\end{center}
\end{figure}

As any piece of electronics equipment, CEA bolometer arrays are susceptible to electromagnetic perturbations. Dedicated tests during the ground calibration campaign of the PACS Photometer shown erratic interferences on the data leading to a wavy pattern in the final projected map when exposed to strong electromagnetic perturbations (see figure~\ref{fig:interf}).
In flight, these interferences are intermittent and short-lived, therefore they usually alter a small fraction of an observation. However these interferences can noticeably degrade the performance of the detectors. Their amplitude is variable, from faint to severe, and affect only the Blue BFP. These perturbations are possibly triggered by the switching on and off of one of the solar panels of the spacecraft, but our analysis of the phenomenon is still under investigation. 

\section{conclusions} 
\label{sect:conclu}

CEA bolometer arrays are performing in space as well as during the ground calibration campaign of the PACS Photometer. After one year in orbit, no performance degradation was observed. There remain minor difficulties to overcome, such as correcting for electrical cross-talk and understanding the origin of interferences, but the overall quality of the data produced by these detectors is of very good quality. 


\acknowledgments     
The authors would like to thank the CEA-Saclay/SAp Bolometer Group and the Herschel/PACS ICC for a fantastic instrument and their help in obtaining and interpreting the data presented in this paper.


\bibliography{pacs_firstYear.bib}   
\bibliographystyle{spiebib}   

\end{document}